\documentstyle{amsppt}
\newcount\refcount
\advance\refcount 1
\def\newref#1{\xdef#1{\the\refcount}\advance\refcount 1}
\newref\brauer
\newref\distII
\newref\unitaryenum
\newref\shorlaflamme

\def\Tr{\operatorname{Tr}}
\def\tensor{\otimes}
\topmatter
\title Polynomial invariants of quantum codes \endtitle
\author Eric M. Rains\endauthor
\affil AT\&T Research \endaffil
\address AT\&T Research, Room 2D-147, 600 Mountain Ave.
         Murray Hill, NJ 07974, USA \endaddress
\email rains\@research.att.com \endemail
\date April 3, 1997\enddate
\abstract
The weight enumerators (\cite\shorlaflamme) of a quantum code are quite
powerful tools for exploring its structure.  As the weight enumerators are
quadratic invariants of the code, this suggests the consideration of
higher-degree polynomial invariants.  We show that the space of degree $k$
invariants of a code of length $n$ is spanned by a set of basic invariants
in one-to-one correspondence with $S_k^n$.  We then present a number of
equations and inequalities in these invariants; in particular, we give
a higher-order generalization of the shadow enumerator of a code, and
prove that its coefficients are nonnegative.  We also prove that the
quartic invariants of a $((4,4,2))$ are uniquely determined, an important
step in a proof that any $((4,4,2))$ is additive (\cite\distII).

\endabstract
\endtopmatter
In \cite\shorlaflamme, Shor and Laflamme introduced the concept of the
weight enumerator of a quantum code, in order to prove a bound on 
the minimum distance of a code, given its length and dimension.  The
weight enumerators have the following two properties: equivalent
codes have equal weight enumerators, and the coefficients of the weight
enumerators are quadratic functions of the projection matrix associated to
the code.  More concisely, we can say that the coefficients of the
weight enumerators are quadratic invariants of the code.  In the present
work, we will consider more general polynomial invariants.

The first task in the exploration is, naturally, to give a characterization
of all polynomial invariants.  Clearly, the invariants of any fixed degree
form a vector space, so it suffices to give a set of invariants that span
that space.  This role is played by what we will call {\it basic}
invariants; as we shall see, the basic invariants of order $k$ and length
$n$ are in one-to-one correspondence with the group $S_k^n$.  In the case
of quadratic invariants, we recover the unitary weight enumerators of
\cite\unitaryenum.

In \cite\unitaryenum, a conjecture is made regarding the {\it shadow
enumerator} of a quantum code, in the case of alphabet size greater than 2.
It turns out that this has a natural generalization to higher-order
invariants; moreover, the structure of the generalization suggests a
natural proof, thus settling that conjecture, and strengthening the linear
programming bound for non-binary quantum codes.

The number of basic invariants is prohibitively large (${k!}^n$) for $n$
and $k$ of any size.  In order to render these invariants tractable, it is
thus necessary that a number of linear dependences be found between them.
In particular, it turns out that in a number of cases, an invariant of
order $k$ can be shown to be equal to an invariant of order $k-1$.  In
addition, if the order of the invariant is greater than the alphabet size,
we get further relations.  In some cases, this reduces the degrees of
freedom to the point of tractability.

We examine how these relations can be used in the case when the quantum
code is a $((4,4,2))$; that is, when the code encodes 4 states in 4
qubits, with minimum distance 2.  In this case, the available relations
allow one to reduce the 331776 basic quartic invariants down to six degrees
of freedom, which can be determined using more ad-hoc methods.  In
particular, we conclude that any two $((4,4,2))$s must have the same
quartic invariants.  In \cite\distII, this fact is used to prove that
any $((4,4,2))$ is equivalent to an additive code, and similarly for
any $((5,2,3))$ or $((6,1,4))$, proving the uniqueness of each of those
codes.

\head Basic polynomial invariants \endhead

For our purposes, it will be convenient to consider two types of polynomial
invariants.  Let $Q$ be a quantum code of length $n$, dimension $K$, and
alphabet size $\alpha$; let $P_Q$ be the associated projection operator.  A
local polynomial invariant of $Q$ is defined as a polynomial function $f$ of
the coefficients of $P_Q$ such that
$$
f(\phi P_Q \phi^{-1})=f(P_Q)
$$
for any $\phi\in U(\alpha)^{\tensor n}$.  A global invariant is then
defined as a local invariant that is also left unchanged under arbitrary
permutations of the letters of $Q$.  While global invariants are the only
{\it true} invariants of the code, the structure of local invariants is
simpler, and determines the global structure; we will therefore begin by
considering local polynomial invariants.

Any polynomial function $f(M)$ in the coefficients of a matrix $M$ can be
written in the following form:
$$
\sum_k \Tr(F^{(k)} M^{\tensor k}),
$$
for a suitable set of matrices $F^{(k)}$ on the domain of $M^{\tensor k}$.
This can be seen by noting that a monomial of degree $k$ in the coefficents
of $M$ can be written as
$$
\prod_{1\le i\le n} \Tr(E_i M),
$$
where each $E_i$ has exactly one coefficent nonzero.  But this is the same
as
$$
\Tr((E_i)^{\tensor k} M^{\tensor k}).
$$
Summing over monomials and over $k$, we get the desired expression.
Now, consider how this expression changes when we conjugate $M$ by a local
equivalence $\phi$:
$$
\sum_k \Tr(F^{(k)} \phi^{\tensor k} M^{\tensor k}(\phi^{\tensor k})^{-1})
=
\sum_k \Tr((\phi^{\tensor k})^{-1} F^{(k)} \phi^{\tensor k} M^{\tensor k}).
$$
In particular, we can average over all local equivalences (since the group
$U(\alpha)^{\tensor n}$ is compact) to obtain a polynomial invariant, and any
polynomial invariant can be taken of that form.  But this is equivalent to
requiring that
$$
F^{(k)}=
(\phi^{\tensor k})^{-1} F^{(k)} \phi^{\tensor k}
\tag 1
$$
for all $\phi\in U(\alpha)^{\tensor n}$.
Thus we have reduced our classification problem to that of classifying
the matrices $F^{(k)}$ satisfying \thetag{1}.

Suppose, for a moment, that $n$ is 1, so the group of local equivalences is
the entire unitary group.  In this case, the classical theory of invariants
of the classical groups (see, for instance, \cite\brauer) tells us that the
space of invariant $F^{(k)}$s is spanned by a set of basic invariants in
one-to-one correspondence with the symmetric group $S_k$.  To be precise,
for $\pi\in S_k$, the corresponding basic invariant is
$$
T(\pi)=
\delta^{i_1}_{i_{\pi(1)}}
\delta^{i_2}_{i_{\pi(2)}}
\delta^{i_3}_{i_{\pi(3)}}
\ldots
\delta^{i_k}_{i_{\pi(k)}}.
$$
Alternatively, if we consider $({\Bbb C}^{\,\alpha\!})^{\tensor k}$ as a
$k$-letter Hilbert space, $T(\pi)$ is the operator which permutes the $k$
qubits according to the permutation $\pi$.  For $n>1$, we can simply remark
that the space of degree $k$ invariants of a tensor product of groups is
equal to the tensor product of the invariant spaces associated to each
group individually.  In particular, this gives us basic invariants in
one-to-one correspondence with $S^n_k$.  The corresponding operators
$T(\pi)$ for $\pi\in S^n_k$ act on $(({\Bbb C}^{\,\alpha\!})^{\tensor
n})^{\tensor k}$ by permuting the $k$ copies of the $i$th qubit according
to the $i$th permutation in $\pi$.

\proclaim{Theorem 1}
Let $f(Q)$ be a polynomial invariant of quantum codes of length $n$ and
 alphabet size $\alpha$.  Then there exists a sequence of functions $f_k$
 on $S^n_k$, eventually zero, such that
$$
f(Q)=\sum_{0\le k} \sum_{\pi\in S^n_k} f_k(\pi) \Tr(T(\pi) (P_Q)^{\tensor
k})
$$
for all codes $Q$ in the domain of $f$.  If $f$ is a global invariant, then
$f_k(\pi)$ can be taken to be invariant under arbitrary permutations of
the $n$ subpermutations of $\pi$.
\endproclaim

\demo{Proof}
The above discussion has proved everything except for the comment on global
invariants; this follows easily from considering the effect that reordering
the qubits has on the basic invariants.
\qed\enddemo

When $k=2$, we recover the unitary weight enumerator of \cite\unitaryenum.
To be precise, note that $S^n_2$ is in one-to-one correspondence with
the set of subsets of $\{1,2,\ldots n\}$; to $\pi\in S^n_2$, we associate
the set $S(\pi)$ of indices such that $\pi_i$ is a transposition.
Then, using the notation of \cite\unitaryenum,

\proclaim{Theorem 2}
For any $\pi\in S^n_2$, and any Hermitian operators $M$ and $N$,
$$
\Tr(T(\pi) (M\tensor N))
=
A'_{S(\pi)}(M,N).
$$
\endproclaim

\demo{Proof}
Recall that $A'_S(M,N)$ is defined by
$$
A'_S(M,N)=\Tr(\Tr_{S^c}(M)\Tr_{S^c}(N)).
$$
Now, for all $i$ in $S^c$, we readily see that
$$
\Tr(T(\pi) (M\tensor N))
=
\Tr(T(\pi^{(i)}) (\Tr_i(M)\tensor \Tr_i(N))),
$$
where $\pi^{(i)}$ is the tuple of permutations obtained from $\pi$ by
removing the $i$th permutation.  This follows, for example, by noting that
we can conjugate $M$ by an arbitrary unitary operation on the $i$th qubit
without changing the result.  But then we get
$$
\Tr(T(\pi) (M\tensor N))
=
\Tr(T((12)) (\Tr_{S^c}(M)\tensor \Tr_{S^c}(N)))
=
A'_S(M,N).
$$
\qed\enddemo

This motivates the notation
$$
A'_\pi(Q)=\Tr(T(\pi) P^{\tensor k}_Q).
$$
Similarly, if we are given a $k$-tuple of Hermitian matrices, we write
$$
A'_\pi(M_1,M_2,\ldots M_k)
=
\Tr(T(\pi) (M_1\tensor M_2\tensor\cdots M_k)).
$$

\head Generalized shadow inequalities \endhead

In \cite\unitaryenum, the author made the following conjecture:

\proclaim{Conjecture} Let $V=V_1\otimes V_2\otimes\cdots\otimes V_n$,
where $V_1$ through $V_n$ are Hilbert spaces.  Let $T$ be any 
subset of $\{1,2,\ldots n\}$, and let $M$ and $N$ be positive
semi-definite Hermitian operators on $V$.  Then
$$
\sum_{S\subset \{1,2,\ldots n\}} (-1)^{|S\cap T|}
\Tr(\Tr_{S^c}(M)\Tr_{S^c}(N))
\ge 0.
$$
\endproclaim

In particular, when $M=N=P_Q$, this gives an inequality that the quadratic
invariants of a code must satisfy.  It turns out that this generalizes
naturally to higher order invariants.  Let us first restate this conjecture
in terms of basic quadratic invariants:
$$
\align
\kappa_T(M,N)&=
\sum_{\pi\in S^n_2} (-1)^{|S(\pi)\cap T|}
A'_\pi(M,N)\\
&=
\sum_{\pi\in S^n_2} \lambda_T(\pi) A'_\pi(M,N).\\
&\ge 0,
\endalign
$$
where
$$
\lambda_T(\pi)=(-1)^{|S(\pi)\cap T|}.
$$
Note that $\lambda_T(\pi)$ is a Hermitian idempotent in the group algebra
of $S^n_2$; this suggests the following generalization:

\proclaim{Theorem 3}
Let $V=V_1\otimes V_2\otimes\cdots\otimes V_n$, where $V_1$ through $V_n$
are Hilbert spaces.  Let $T$ be any subset of $\{1,2,\ldots n\}$, and let
$M_1$, $M_2$,\dots $M_k$ be positive semi-definite Hermitian operators on
$V$.  Then for any Hermitian idempotent $\lambda$ in the group algebra of
$S^n_k$,
$$
\sum_{\pi\in S^n_k} \lambda(\pi) A'_\pi(M_1,M_2,\dots M_k)\ge 0.
$$
\endproclaim

\demo{Proof}
The given expression is multilinear in the $M_i$; consequently, it suffices
to consider the case in which each $M_i$ has rank one.  Thus, let $v_i$
for each $i$ be a vector in $V$ such that $M_i=v_i v^\dagger_i$.  This
allows us to restate our question as showing that:
$$
\kappa=\sum_{\pi\in S^n_k} \lambda(\pi) A'_\pi(v_i v^\dagger_i)\ge 0.
$$
Rewriting $A'_\pi$ as a trace, we have:
$$
\align
\kappa&=\sum_{\pi\in S^n_k} \lambda(\pi) 
(v_1\tensor v_2\tensor \ldots v_k)^\dagger
T(\pi)
(v_1\tensor v_2\tensor \ldots v_k)\\
&=
\sum_{\pi,\pi'\in S^n_k} \lambda(\pi\pi^{\prime{-}1})\lambda(\pi')
(v_1\tensor v_2\tensor \ldots v_k)^\dagger
T(\pi)
(v_1\tensor v_2\tensor \ldots v_k)\\
&=
\sum_{\pi_1,\pi_2\in S^n_k} \lambda(\pi_1)\lambda(\pi_2)
(v_1\tensor v_2\tensor \ldots v_k)^\dagger
T(\pi_1) T(\pi_2)
(v_1\tensor v_2\tensor \ldots v_k)\\
&=
(\sum_{\pi_1\in S^n_k} \lambda(\pi_1) T(\pi_1) v_1\tensor v_2\tensor \ldots v_k)^\dagger
(\sum_{\pi_2\in S^n_k} \lambda(\pi_2) T(\pi_2) v_1\tensor v_2\tensor \ldots v_k).
\endalign
$$
But this is the norm of a vector, so must be nonnegative.
\qed\enddemo

By the discussion preceding the theorem, the above conjecture follows as
an immediate corollary.

\head Reductions and relations\endhead

Although we have shown that the basic invariants span the space of
polynomial invariants, we have by no means shown that they form a basis.
Indeed, there are a number of linear equations relating the various basic
invariants.  For example, if we conjugate every permutation in $\pi$ by a
fixed element of $S_k$, the corresponding basic invariant will be
unchanged; this corresponds to the fact that the different copies of $P_Q$
appearing in the expression for the basic invariant can be freely
interchanged.

Many of these equations take the form of a reduction, in which an
invariant of degree $k$ is expressed as an invariant of degree $k-1$.
The most general of these reductions follows from the fact that $P_Q$ 
is a projection operator, so $P^2_Q=P_Q$.  Suppose there is an index
$1\le j\le n$ such that $\pi_i(j)$ is constant as $i$ ranges from
1 to $k$.  Then each qubit of the $j$th copy of $P_Q$ is connected to the
corresponding qubit of the $\pi_i(j)$th copy.  But then this gives us
$P^2_Q$, which we can replace by $P_Q$.

For example, consider the invariant
$$
A'_{(12)(345),(123)(45),(124)(35),(125)(34)}(P_Q).\tag{2}
$$
Here, each permutation maps $1$ to $2$; in consequence, we can merge $1$
and $2$, obtaining:
$$
A'_{(1)(345),(13)(45),(14)(35),(15)(34)}(P_Q),
$$
which we can renumber as
$$
A'_{(1)(234),(12)(34),(13)(24),(14)(23)}(P_Q),
$$
which we are unable to reduce further.

In some cases, the information we are given concerning $Q$ allows us to
give further reductions.  For example, suppose we are given a set $S$ such
that $A'_S(Q)=2^{-|S|} A'_{\emptyset}(Q)$; if $Q$ is known to be pure to
weight $w$, for instance, then this holds for all $S$ of size less than
$w$.  Then $\Tr_{S^c}(P_Q)$ is proportional to an identity matrix (with
some known constant depending only on the dimension of $Q$).  Suppose there
exists a $j$ with $1\le j\le k$ such that $\pi_i(j)=j$ for all $i$ in
$S^c$.  Then we can splice $j$ out of each permutation, resulting in a
lower-order invariant that is a constant multiple of the original
invariant.

For instance, suppose that in the above example we knew that
$\Tr_{S^c}(P_Q)={1\over 2} I$ for all $S$ of cardinality 3; this is the
case when $Q$ is a binary $((4,4,2))$, for instance.  Then we can reduce
$$
A'_{(1)(234),(12)(34),(13)(24),(14)(23)}(P_Q)
$$
by splicing out 1, obtaining
$$
{1\over 2} A'_{(234),(2)(34),(3)(24),(4)(23)}(P_Q),
$$
which reduces further to
$$
{1\over 4} A'_{(34),(34),(3)(4),(3)(4)}(P_Q) = {1\over 4} A'_{\{1,2\}}(P_Q).
$$
Thus we have reduced the quintic invariant \thetag{2} to a quadratic
invariant.  It should be apparent, therefore, that these reductions can be
a powerful tool in simplifying higher-order invariants.

A final important class of relations appears when the order of the
invariant is greater than the alphabet size.  For $n=1$, we have the
following classical result:

\proclaim{Lemma 4}
If $k>\alpha$, then
$$
\sum_{\pi\in S_k} \sigma(\pi) T(\pi)=0.
$$
\endproclaim

\demo{Proof}
Let $M$ be a matrix of dimension $k$.  Then one readily sees that
$$
\Tr((\sum_{\pi\in S_k} \sigma(\pi) T(\pi))
    M^{\tensor k})
=
\det(M),
$$
from the definition of determinant.  Now, the basic invariants are
unchanged if we enlarge each matrix by adding a row and column of zeros.
Consequently, for $M$ of dimension less than $k$ (i.e., $\alpha$),
$$
\Tr((\sum_{\pi\in S_k} \sigma(\pi) T(\pi))
    M^{\tensor k})
=
0.
$$
The only way this can happen for all $M$ is if
$$
\sum_{\pi\in S_k} \sigma(\pi) T(\pi)=0.
$$
\qed\enddemo

We get further relations by adding fixed points, or multiplying the sum of
$T$s by some fixed $T$.  For instance, consider the invariant
$$
A'_{(123),(123),(132)}(Q),
$$
in the case when $Q$ is a binary code.  Then lemma 4 tells us
that
$$
\multline
A'_{(123),(123),(132)}(Q)
+
A'_{(123),(123),(1)(2)(3)}(Q)
+
A'_{(123),(123),(123)}(Q)\\
=
A'_{(123),(123),(12)(3)}(Q)+
A'_{(123),(123),(13)(2)}(Q)+
A'_{(123),(123),(23)(1)}(Q).
\endmultline
$$
The last four terms always reduce to quadratic invariants, while the second
term sometimes admits reduction as well.  We also have, for example,
$$
\multline
A'_{(12)(34),(123)(4)}(Q)+A'_{(134)(2),(123)(4)}(Q)+A'_{(234)(1),(123)(4)}(Q)
\\=
A'_{(1)(2)(34),(123)(4)}(Q)+A'_{(1342),(123)(4)}(Q)+A'_{(1234),(123)(4)}(Q),
\endmultline
$$
obtained by multiplying the relation
$$
\multline
T((1)(2)(3)(4))+T((123)(4))+T((132)(4))
\\=
T((12)(3)(4))+T((13)(2)(4))+T((1)(23)(4))
\endmultline
$$
by $T((12)(34))$.  As we shall see in the next section, these relations
derived from lemma 4 can be extremely powerful.

\head Binary MDS codes of distance 2 \endhead

Let us consider the case when $Q$ is a binary MDS code of distance 2; that is,
when $Q$ is a $((2m,4^{m-1},2))$.  In this case, $\Tr_i(P_Q)={1\over 2} I$
for $1\le i\le 2m$.  This allows us to reduce any basic invariant
containing a non-derangement (a permutation with a fixed point) to a
lower-order invariant, as we have just seen.

\proclaim{Lemma 5}
For each $m\ge 2$, the local cubic invariants of a $((2m,4^{m-1},2))$ are
uniquely determined.  That is, if $Q$ and $Q'$ are $((2m,4^{m-1},2))$s, and
$\pi\in S_3^{2m}$, then $A'_\pi(Q)=A'_\pi(Q')$.
\endproclaim

\demo{Proof} Since a $((2m,4^{m-1},2))$ is MDS, its local quadratic
invariants are uniquely determined.  Therefore, it suffices to restrict our
attention to those invariants corresponding to $\pi$ consisting entirely of
derangements; otherwise, the invariant can be reduced to a local quadratic
invariant, and is thus uniquely determined.

Since we are dealing with a binary code, lemma 4 applies:
$$
T((1,3,2))=-T((1)(2)(3))-T((1,2,3))+T((1,2)(3))+T((1,3)(2))+T((2,3)(1)).
$$
In particular, this allows us to reduce any invariant involving $(1,3,2)$
to a sum of invariants involving only $(1,2,3)$ or permutations with
fixed points.  Thus the only remaining non-derangement invariant is
$
A'_{(1,2,3)^{2m}}(Q),
$
which reduces to
$
\Tr(P_Q^3) = 2^{m-2}.
$
\qed\enddemo

For a $((4,4,2))$, we can say more:

\proclaim{Theorem 6}
The local quartic invariants of a $((4,4,2))$ are uniquely determined.
\endproclaim

\demo{Proof}
Let $Q$ be a $((4,4,2))$.  As in lemma 5, we may restrict our attention to
derangements.  For convenience, we define
$$
\pi_1=(1,2,3,4),\quad \pi_2=(1,3,4,2), \quad \pi_3=(1,4,2,3).
$$
Then any derangement in $S_4$ can be written as $\pi_i^j$ for $i,j\in
\{1,2,3\}$.  Furthermore, we have the following relations in the
representation $T$, from lemma 4:
$$
\matrix
\pi_1^3=\pi_1+\text{n.d.},\quad \pi_2^3=\pi_2+\text{n.d.}, \quad
\pi_3^3=\pi_3+\text{n.d.}\\
\pi_1^2=\pi_2+\pi_3+\text{n.d.}, \quad
\pi_2^2=\pi_1+\pi_3+\text{n.d.}, \quad
\pi_3^2=\pi_1+\pi_2+\text{n.d.},
\endmatrix
\tag 3
$$
where n.d. refers to some linear combination of non-derangements.  This
allows us to restrict our attention to invariants involving only $\pi_1$,
$\pi_2$, and $\pi_3$.  Now, note that $\pi_1(3)=\pi_2(3)$,
$\pi_2(4)=\pi_3(4)$, and $\pi_1(2)=\pi_3(2)$.  It follows that any
invariant involving only two of the three can be reduced to a cubic
invariant.  We therefore have only six degrees of freedom remaining,
corresponding to the local invariant
$$
A'_{\pi_1,\pi_1,\pi_2,\pi_3}(Q),
$$
and its six permutations.  In particular, we have only one degree of
freedom remaining in the global invariants.

Now, let $v$ be any codeword in $Q$, and consider $\Tr_{\{234\}}(vv^\dagger)$.
Since $Q$ is pure to distance 2, it follows that
$$
\Tr_{\{234\}}(v v^\dagger)=2 I.
$$
Thus the commutator
$$
[\Tr_{\{234\}}(v v^\dagger)\tensor I,P_Q]=0
$$
for all $v\in Q$.  But then
$$
E_{v\in Q}(\Tr(-[\Tr_{\{234\}}(v v^\dagger)\tensor I,P_Q]^2))=0,
$$
where $v$ may be taken to be normally distributed.
This is a local quartic invariant of $Q$, equal to
$$
\align
 2 &A'_{(1,2,3,4),(1)(2)(3,4),(1)(2)(3,4),(1)(2)(3,4)}(Q)\\
{}+2 &A'_{(1)(2,3,4),(1,2)(3,4),(1,2)(3,4),(1,2)(3,4)}(Q)\\
{}-2 &A'_{(1,3,2,4),(1)(2)(3,4),(1)(2)(3,4),(1)(2)(3,4)}(Q)\\
{}-2 &A'_{(1,3)(2,4),(1,2)(3,4),(1,2)(3,4),(1,2)(3,4)}(Q).
\endalign
$$
Simplifying along fixed points, we conclude that
$$
A'_{(1,3)(2,4),(1,2)(3,4),(1,2)(3,4),(1,2)(3,4)}(Q)=
A'_{\pi_1^2,\pi_3^2,\pi_3^2,\pi_3^2}=
4.
$$
Applying the reductions \thetag{3}to the first qubit, we find
$$
A'_{\pi_3,\pi_3^2,\pi_3^2,\pi_3^2}(Q)=2,
$$
then, reducing the second through fourth qubits,
$$
A'_{\pi_3,\pi_1,\pi_3^2,\pi_3^2}(Q)+
A'_{\pi_3,\pi_2,\pi_3^2,\pi_3^2}(Q)=2,
$$
$$
A'_{\pi_3,\pi_1,\pi_1,\pi_3^2}(Q)+
A'_{\pi_3,\pi_2,\pi_1,\pi_3^2}(Q)+
A'_{\pi_3,\pi_1,\pi_2,\pi_3^2}(Q)+
A'_{\pi_3,\pi_2,\pi_2,\pi_3^2}(Q)
=2,
$$
and
$$
A'_{\pi_3,\pi_1,\pi_1,\pi_2}+
A'_{\pi_3,\pi_1,\pi_2,\pi_1}+
A'_{\pi_3,\pi_2,\pi_1,\pi_1}
=
3.
$$
Permuting this equation gives us four more equations relating the local
invariants, leaving two degrees of freedom.  However, this is enough to
determine the global invariants:
$$
(A'_{\pi_1,\pi_1,\pi_2,\pi_3})_{\text{sym}} = 1.
$$
Now, consider, for $S\subset \{1,2,3,4\}$ of size 2,
$$
\kappa_S = E_{v\in Q}(\Tr(-[\Tr_{S^c}(v v^\dagger)\tensor I,P_Q]^2)).
$$
This is a local quartic invariant, and further must be nonnegative,
since the commutator of two Hermitian operators is anti-Hermitian.
On the other hand, we have, for example,
$$
\align
\kappa_{\{1,2\}}\propto{}
&
8
-2 A'_{\pi_2,\pi_3,\pi_1,\pi_1}(Q)
-2 A'_{\pi_1,\pi_1,\pi_2,\pi_3}(Q)\\
&{}-A'_{\pi_1,\pi_2,\pi_1,\pi_3}(Q)
-A'_{\pi_1,\pi_2,\pi_3,\pi_1}(Q)
-A'_{\pi_2,\pi_1,\pi_1,\pi_3}(Q)
-A'_{\pi_2,\pi_1,\pi_3,\pi_1}(Q)
.
\endalign
$$
Symmetrizing, we find
$$
\sum_S \kappa_S = 0.
$$
But then the nonnegativity of the $\kappa_S$ implies $\kappa_S=0$ for each
$S$.  This gives us three further equations on the local invariants,
eliminating the two remaining degrees of freedom.
\qed\enddemo

\head Further directions\endhead

It must be stressed that the relations and inequalities we have derived
above by no means exhaust the possibilities; for instance, it should be
possible to define a higher-order, but still nonnegative, analogue of the
$A_S$ weight enumerators, which would produce a number of inequalities
on the higher-order invariants.  Furthermore, it seems clear that we are
not taking full advantage of the relations that can be deduced from the
minimum distance of the code.  Progress needs to be made in these areas
in order for polynomial invariants to be truly useful.

In addition, it should be noted in passing that there is some evidence (see
the remark after corollary 9 in \cite\distII) that some simple set of
relations on the quartic invariants, satisfied by all additive codes, are
enough to force a code to be additive.  This possibility probably merits
further investigation.

\enddocument
\bye